\documentclass[a4paper,10pt]{IEEEtran}
\usepackage[latin1]{inputenc}
\usepackage{graphicx}
\usepackage{amsmath}
\usepackage{amssymb}
\usepackage{pslatex}
\title{An Analytical Model of the IEEE 802.3ah MAC Protocol for EPON-based Access
Systems\bigskip}
\author{\authorblockN{Sawsan Al Zahr and Maurice Gagnaire}\\
\authorblockA{Department of Computer Sciences and Networks\\
Telecom Paris - LTCI - UMR 5141 CNRS\\
46, rue Barrault F 75634 Paris - France\\
Email: \{sawsan.alzahr,maurice.gagnaire\}@enst.fr}}
\begin{document}
\maketitle
%=================================================================
\begin{abstract}
Ethernet Passive Optical Networks (EPON) access systems are
considered as an alternative to ADSL for high speed access to the
Internet. Supported by the Ethernet First Mile Alliance and the
IEEE, the MPCP protocol has been adopted as the standardized MAC
protocol for EPONs. Moreover, the IPACT dynamic bandwidth allocation
mechanism has been proposed for hierarchical multiplexing. Today,
IPACT is considered as interesting complement to MPCP for dynamic
bandwidth allocation over EPONs. In this paper, we propose an
original analytical model of the MPCP+IPACT protocol. Numerical
results obtained from this model are commented and compared to
computer simulations.
\end{abstract}
%=================================================================
%=================================================================
\section{Introduction}
%=================================================================
Many investigations have been carried these last fifteen years to
evaluate the feasibility of fiber-in-the-loop (FITL) access systems.
During the years $1990$, several major telecommunication carriers
associated in the FSAN (Full Service Access Network) initiative have
specified the concept of PON (Passive Optical Network). A PON is a
passive point-to-multipoint optical access system. Two types of
modems are used in a PON: the Optical Line Termination (OLT) at the
head of the optical tree, and the Optical Network Units (ONUs) at
the customer premises or close to these premises. For economy
purposes, only two optical wavelengths are used in the system. A
common upstream wavelength at $1300$ nm is used at the ONUs whereas
the OLT uses another wavelength at $1550$ nm for downstream traffic.
Standardized by the ITU-T (G.$983$), such systems are designed for
the transport of ATM connections. The principle of APONs is now
extended for Wavelength Division Multiplexing (WDM) and protection
and restoration with the BPONs (Broadband PON) and GPONs (Gigabit
PON) concepts. With the emergence of Ethernet switching in the
metropolitan area since the year $2000$, the IEEE and the Ethernet
First Mile Alliance are promoting the concept of Ethernet PON (EPON)
also known as the IEEE 802.3ah standard. Physical layer
functionalities are roughly the same in APONs and EPONs. Both APONs
and EPONs require a Medium Access Control (MAC) protocol in order to
prevent collisions when MAC-PDUs are transmitted from the ONUs. The
MPCP MAC protocol has been developed by the IEEE 802.3ah Task Force
\cite{MPCP:IEEE}. It is based on the concept of non-overlapping
upstream transmission windows (timeslots) allocated to each ONU by
the OLT. In addition to MPCP, multiple Dynamic Bandwidth Allocation
(DBA) schemes have been proposed these last three years as
complement of MPCP \cite{Kramer:FQSE, Ma:2003}. These DBA schemes
aim either to efficient statistical multiplexing or to Quality of
Service (QoS) provisioning in EPON systems \cite{McGarry:2004}. In
this paper, we propose the first (at our knowledge) analytical model
of the MPCP+IPACT protocol. In section II, we recall the main
characteristics of the IEEE 802.3ah MAC protocol and of the IPACT
dynamic bandwidth allocation scheme. In section III, we describe our
analytical model of MPCP+IPACT. Finally, we present and comment in
section IV numerical results obtained by means of this analytical
model.

%=================================================================
\section{MAC Protocol and Bandwidth Allocation}
%=================================================================
In this section, we describe successively the principle of the IEEE
802.3ah MAC protocol and of the IPACT DBA mechanism.
%=================================================================
\subsection{Mutipoint Control Protocol (MPCP) Overview}
%=================================================================
Although MPCP is not concerned with any particular bandwidth
allocation (QoS dependant), it has been designed in order to
facilitate the implementation of various DBA algorithms. MPCP relies
on two Ethernet control messages (GATE and REPORT) to provide the
signalling infrastructure (control plane) for coordinating upstream
data transmission. Three other signalling messages are used for
automatic ONUs discovery (REGISTER\_REQUEST, REGISTER, and
REGISTER\_ACK). Such messages enable for instance distance
equalization between the OLT and the various ONUs.

In its regular operation, the OLT generates downstream GATE messages
to dynamically assign timeslots to the active ONUs. A GATE message
contains the time when an ONU is allowed to start its transmission
and the length of this transmission window (timeslot). It is the
role of each ONU to coordinate access to the medium during its
allocated timeslot among its different active traffic queues. In
order to facilitate this coordination, MPCP stamps each GET message
when it arrives at an ONU with a local clock. Upon receiving a
message matching its MAC address, each ONU updates its local clock
to that of the timestamp in the received GATE control message to
avoid any potential clock drift. A transmission window may include
multiple variable size Ethernet frames. Depending on the size of the
allocated timeslot and the number of buffered packets (Ethernet
frames) at the ONU, a REPORT message followed by upstream user data
frames is sent in each allocated timeslot. Typically, a REPORT
message contains the required size of the next timeslot based on the
ONUs buffer occupancy. Upon receiving REPORT messages, the OLT
passes the bandwidth requests of the various ONUs to the DBA module
responsible for bandwidth allocation decision.
%=================================================================
\subsection{Interleaved Polling with Adaptive Cycle Time (IPACT)}
%=================================================================
Interleaved Polling with Adapting Cycle Time (IPACT), proposed in
\cite{Kramer:IPACT}, provides a statistical multiplexing for ONUs
and results in efficient upstream channel utilization. IPACT
consists in a pipelining process using an interleaved polling
strategy. For instance, ONU$_{i}$ is polled during ONU$_{i-1}$ is
sending its user data to the OLT. In order to facilitate this
pipelining, an ONU may generate its own control message which is
piggybacked to user data transmission. Thus, a given ONU informs the
OLT by means of a piggybacked control message how many bytes were in
its buffer at the instant of its last user data transmission. In
doing so, bandwidth is dynamically assigned to ONUs according to
their buffer occupancy. An ONUs which is transitory not allowed to
send data during a given transmission cycle is still polled by the
OLT. This ONU is then able to report its queues occupancy for the
next cycle. By convention, the transmission of a user frame that
cannot fit in totality in a timeslot is postponed to the next
timeslot. To prevent the upstream channel being monopolized by a
single ONU with high data volume, a maximum transmission window size
is assigned to each ONU.

We denote the specific maximum transmission window size of ONU$_{i}$
by $W_{MAX}^{[i]}$ (in bytes). The choice of the specific values of
$W_{MAX}^{[i]}$, assigned to the various ONUs, determines the
maximum polling cycle time $T_{MAX}$ under heavy load conditions:
\begin{equation}\label{eq:tmax}
T_{MAX} = \sum_{i=1}^{N} \left(G + \frac{8 \times
W_{MAX}^{[i]}}{R_U} \right)
\end{equation}
where $G$ is the guard interval (seconds), $N$ is the number of
ONUs, and $R_{U}$ is the line rate (bps). The guard intervals
provide protection for fluctuations of round-trip time (RTT) of
different ONUs (distance ranging). Furthermore, the OLT receiver
needs some time to proceed to two physical layer operations. First,
the OLT needs to readjust its optical power detection thresholds
according to its distance from the transmitting ONU (power ranging).
Second, a clock resynchronization of the PLL (Phase Locked Loop)
must be carried at the OLT at the beginning of each timeslot (clock
recovery). The choice of the $W_{MAX}^{[i]}$ values will also
determines the guaranteed bandwidth available for each ONU$_{i}$. We
denote the minimum guaranteed bandwidth of ONU$_{i}$ by
$\Lambda_{MIN}^{[i]}$ (bps). Obviously, the ONU is guaranteed to be
able to send at least $W_{MAX}^{[i]}$ (bytes) in at most $T_{MAX}$
(seconds):
\begin{equation}\label{eq:miniband}
\Lambda_{MIN}^{[i]} = \frac{8 \times W_{MAX}^{[i]}}{T_{MAX}}
\end{equation}

An ONUs bandwidth is limited to its guaranteed bandwidth only if all
other ONUs in the system also use all of their available bandwidth.
If at least one ONU does not consume its guaranteed bandwidth, it is
assigned a shorter transmission window, thus making the polling
cycle time shorter. Therefore the available bandwidth to all other
ONUs is in this case increased proportionally to their
$W_{MAX}^{[i]}$. As a consequence, the polling cycle time is not
static but is adapted to the instantaneous network load.
%=================================================================
\section{Analitical Model}
%=================================================================
\subsection{Bandwidth management}
%=================================================================
Bandwidth management and fair scheduling of different traffic
classes play very important role in supporting QoS in EPON-based
access network. Diffserv \cite{Diffserv:1998}, developed by Internet
Engineering Task Force (IETF), provides method to classify the
network traffic.

In our model, we classify network traffic into three priorities as
defined in \cite{Diffserv:1998}: the expedited forwarding (EF), the
assured forwarding (AF), and the best effort (BE). EF services
gather the delay sensitive applications that require a bounded
end-to-end delay and jitter specifications (such as voice over IP),
whereas AF class is intended for services that are not delay
sensitive but which require bandwidth guarantees. Finally, BE
services are not delay sensitive and do not require neither jitter
specifications nor minimum guaranteed bandwidth.
\begin{figure}[!h]
\centering
\includegraphics[width=0.45\textwidth]{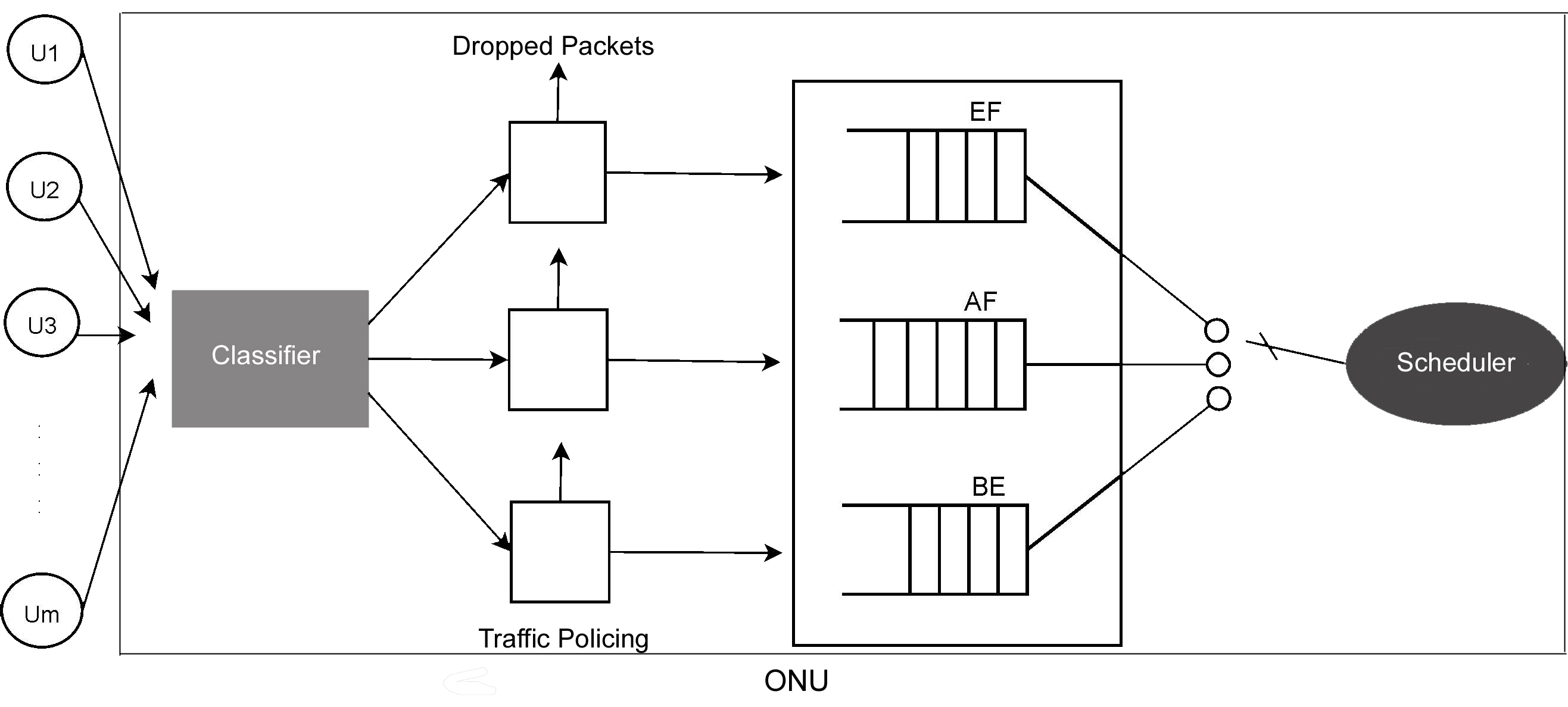}
\caption{ONU queue management} \label{onu-management}
\end{figure}
As shown in Fig. \ref{onu-management}, each ONU is provided with
buffering space shared by three separate priority queues that are
corresponding to service classes. Packets are first classified
according to their type of service and then placed into their
appropriate priority queues. Traffic policing is required at the ONU
to ensure that packets conform to their service level agreement
(SLA), non-conforming traffic being dropped. In the following, we
assume that submitting traffic is conforming. As a result, we may
neglect the traffic policing.

Packets transmission scheduling is performed by a priority-based
scheduler. Strict priority scheduling algorithm, also known as
Head-of-Line (HoL), schedules packets from the head of given queue
only if all higher priority queues are empty. The HoL policy may
then be very unfair for the lowest priority traffic. In order to
prevent this drawback, we propose a priority-based scheduling
mechanism which takes into account the traffic load in addition to
its priority level.
%=================================================================
\subsection{ONU Model}
\begin{figure}[!h]
\centering
\includegraphics[width=0.45\textwidth]{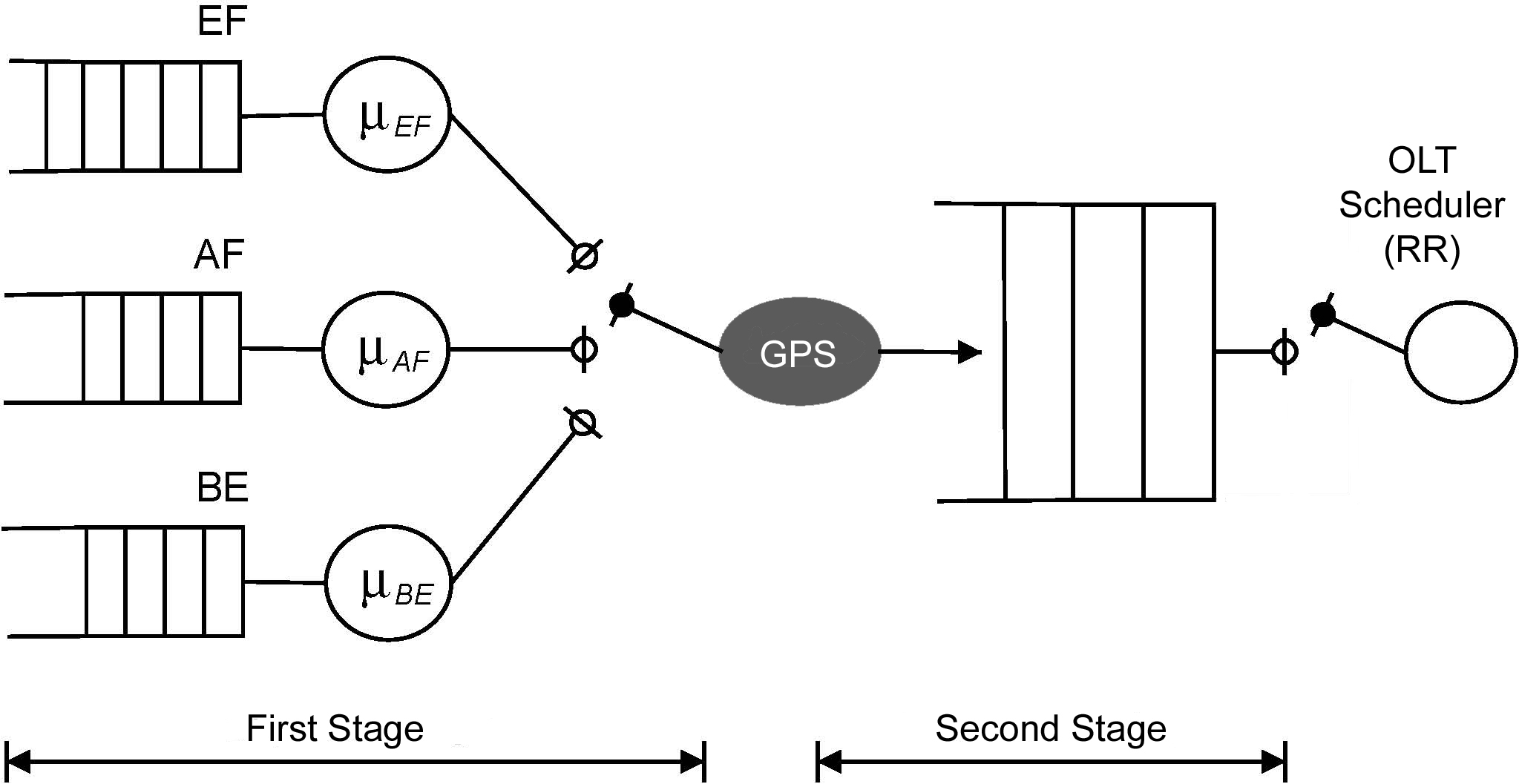}
\caption{ONU model} \label{onu-model1}
\end{figure}

%=================================================================
Our ONU model is shown in Fig. \ref{onu-model1}. It consists of two
separate stages. The first stage is composed of three separate
priority queues as described previously, while the second stage
contains only one common queue. Packets are scheduled in the second
stage according to generalized processor sharing (GPS) strategy.

Each queue has a service share $\varphi_{c}$. In order to fairly
distribute allocated bandwidth, we define the class service share
considering its traffic load:
\begin{equation}
\varphi_{c} = \frac{\lambda_{c}}{\lambda} \times \delta_{c}
\end{equation}
where $\lambda_c$ is the class $c$ traffic load, $\lambda$ is the
global traffic load (i.e. $\lambda = \sum_{c} \lambda_c$), and
$\delta_{c}$ is the weight assigned to the class $c$ based on its
priority, with $\sum_{c} \delta_{c} = 1$. We notice that
$\lambda_{c}$ does not depend on index $i$, i.e. multi-priority
traffic distribution being identical over the different ONUs. GPS
strategy splits allocated bandwidth among all non-empty queues
simultaneously, in proportion to their service shares as follows:
\begin{equation}
\mu_{c}^{[i]} = \frac{\varphi_{c}}{\sum_{c \in Q} \varphi_{c}}
\Lambda_{MIN}^{[i]}
\end{equation}
where $Q$ denotes the set of non-empty queues.
\begin{figure}[!h]
\centering
\includegraphics[width=0.45\textwidth]{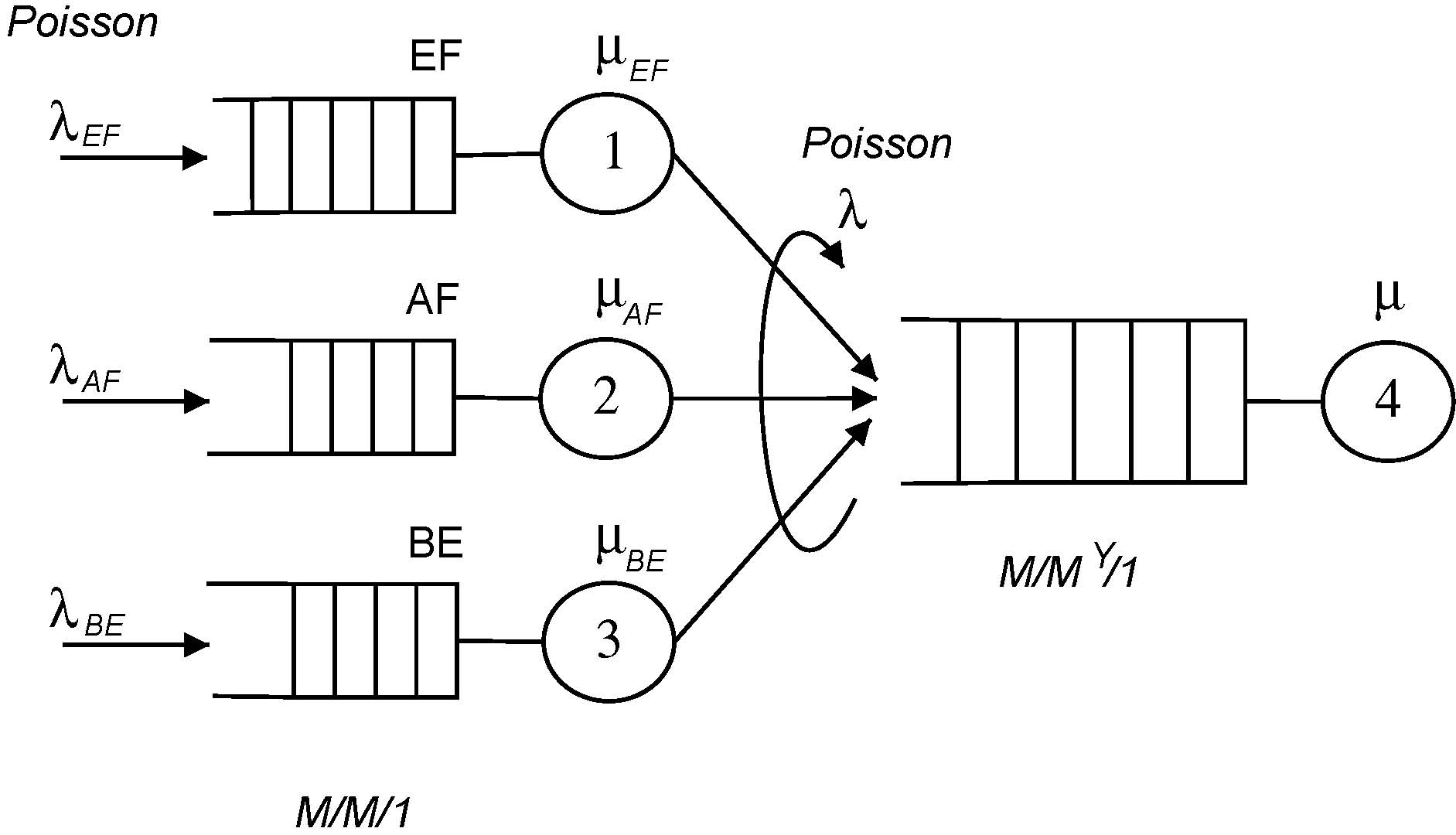}
\caption{Multi-class Open queueing network} \label{onu-model2}
\end{figure}

We assume that new packets arrive at the system according to a
Poisson process with rate $\lambda_{c}$ and have an exponential
service time distribution with mean $1/\mu_{c}^{[i]}$, where $c$
corresponds to traffic class. Since we are mainly interested in the
average access delay and not in packet loss, we can suppose that ONU
buffer is infinite. Thus, each queue of the first stage can be
modeled as a $M/M/1$ queue which has been analyzed extensively in
the literature \cite{Baynat:2000, Pujolle:1989}. For $M/M/1$ queue
at equilibrium, packets departure process will also be Poisson, as
well as combining independent Poisson processes leads to a process
which will also be Poisson in nature \cite{Baynat:2000}.
Furthermore, the average flow rate leaving the $M/M/1$ queue at
equilibrium is the same as the average flow rate entering this
queue. Therefore, packets arrival at the second stage is a Poisson
process with rate $\lambda$. Assume that service time is
exponentially distributed with parameter
$\mu^{[i]}=\Lambda_{MIN}^{[i]}$ and that up to $W_{MAX}^{[i]}$
packets are served simultaneously in considering packets
transmission delay is negligible compared to queueing delay. Hence,
the second stage can be modeled as a $M/M^{Y}/1$ queue, which has
been discussed in \cite{Pujolle:1989}. Fig. \ref{onu-model2} shows
the obtained queueing system which is referred in queueing theory to
multi-class open queueing network.
%=================================================================
\subsection{Performance evaluation}
\begin{figure}\centering
\includegraphics[width=0.45\textwidth]{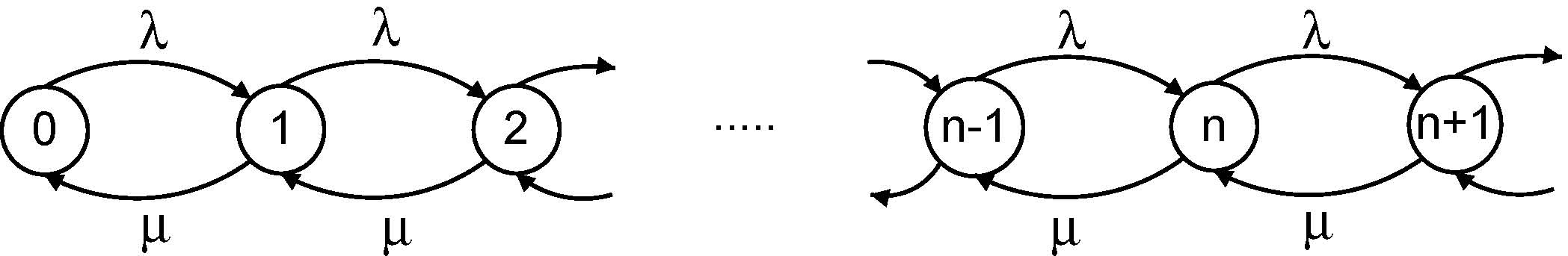}
\caption{Markov chain associated to $M/M/1$ queue}
\label{cmtc-model1}
\end{figure}

\begin{figure}
\centering
\includegraphics[width=0.45\textwidth]{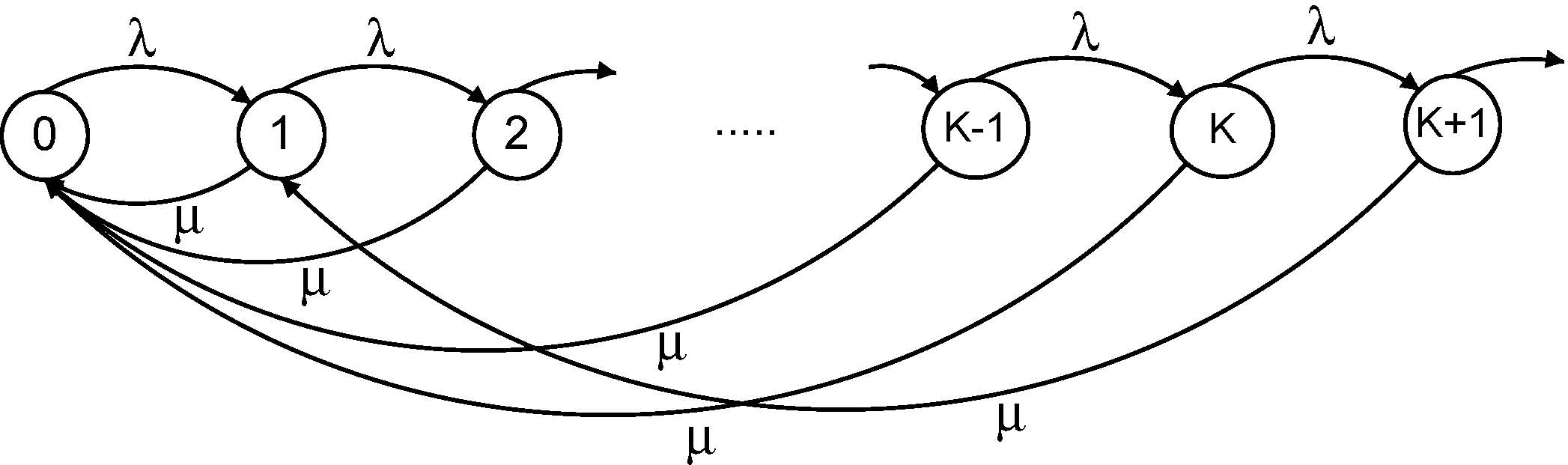}
\caption{Markov chain associated to $M/M^{Y}/1$ queue}
\label{cmtc-model2}
\end{figure}

%=================================================================
In our study, we consider a multi-class open queueing network which
assumes the following hypothesis: (1) it contains only mono-server
stations that operate under FIFO discipline; (2) service time is
exponentially distributed; (3) packets arrival is a Poisson process.
Thus, this network is equivalent to a mono-class open queueing
network, also called open Jackson's network. As a result, the state
probability can be expressed as follows:
\begin{equation}
p(n_{1}, n_{2}, n_{3},n_{4}) = \prod_{i=1}^{4} p_{i}(n_{i})
\end{equation}
where $p_{i}(n_{i})$ are the marginal probabilities associated to
the EF, AF, BE, and $M/M^{Y}/1$ queues respectively. The Markov
chain associated to the three queues of first stage ($i=1$, $i=2$,
and $i=3$) is illustrated by figure \ref{cmtc-model1}. Similarly,
the Markov chain associated to the second stage is illustrated by
figure \ref{cmtc-model2}. As mentioned previously, we are interested
in the average access delay $\text{E}[T]$. In order to determine the
analytical expression of $\text{E}[T]$, we need other performance
parameters, namely the total network throughput $\gamma$ and the
expected number of packets $\text{E}[N]$, where $N$ stands for the
random variable associated to the number of packets in the
system.\medskip

All performance parameters are computed at equilibrium with regard
to queueing network stability. System stability condition is defined
as following:
\begin{equation}
\rho_{i} < 1
\end{equation}
where $\rho_{i} = \lambda_{i}/\mu_{i}$ and $i$ refers the queue.
Given that queues buffer are infinite and all queues are in stable
state, then the total queueing system throughput is:
\begin{equation}\label{eq:throughput}
\gamma = \lambda = \sum_{c} \lambda_{c}
\end{equation}

The expected total number of packets in the queueing system is given
by:
\begin{equation}\label{eq:total-packets-expected-number}
\text{E}[N] = \sum_{i = 1}^{4} \text{E}[N_{i}]
\end{equation}
where $\text{E}[N_{i}]$ is the expected number of packets at queue
$i$. We admitted that all first stage queues are $M/M/1$ queues then
we can compute their expected number of packets as following:
\begin{equation}\label{eq:packets-expected-number1}
\text{E}[N_{i}] = \frac{\rho_{i}}{1 - \rho_{i}}, \quad i \in \{1, 2,
3\}
\end{equation}

Moreover, the second stage which has been modeled with a batch
services queue ($M/M^{Y}/1$) contains a mean number of packets given
by:
\begin{equation}\label{eq:packets-expected-number2}
\text{E}[N_{4}] = \frac{r_{0}}{1 - r_{0}}
\end{equation}
where $r_{0}$ is the single root of Eq. \ref{eq:second-stage-state}
which satisfies $|r_0|<1$, obtained by means of Rouche's theorem
\cite{Pujolle:1989}.
\begin{gather}\label{eq:second-stage-state}
[\mu p^{K+1}(n+1) - (\lambda + \mu)p(n+1)+ \lambda p(n)] = 0
\end{gather}

Finally, we can compute the average access delay using Little's
formula:
\begin{gather}\label{eq:average-access-delay}
\text{E}[T]=\frac{\text{E}[N]}{\gamma}=\frac{r_0}{\lambda(1 - r_0)}
+ \frac{1}{\lambda} \sum_{c} \frac{\rho_{c}}{(1 - \rho_{c})}
\end{gather}
%=================================================================
\section{Numerical Results}
%=================================================================
In this section, we study the impact of priority queueing on the
overall performance of the network and we compare our analytical
results with simulation results presented in \cite{Luo:2005}. We
consider a PON access system with $16$ ONUs and all of them have the
same SLA. The upstream channel capacity $R_{u}$ is equal to $1$
Gbps. The maximum cycle time $T_{MAX}$ is set to $2$ ms and the
guard time $G$ separating two consecutive transmission windows is
set to $5$ $\mu$s. The Ethernet frame size is fixed at $1500$ bytes.
Using both of Eq. \ref{eq:tmax}, and Eq. \ref{eq:miniband} and
taking into account the fact that all the ONUs have the same SLA, we
obtain the maximum size of transmission windows and the minimum
guaranteed bandwidth: $W_{MAX}=15000$ bytes, $\Lambda_{MIN}=60$
Mbps.

\begin{figure}
\centering
\includegraphics[width=0.45\textwidth]{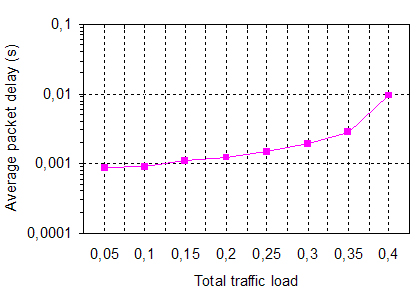}
\includegraphics[width=0.45\textwidth]{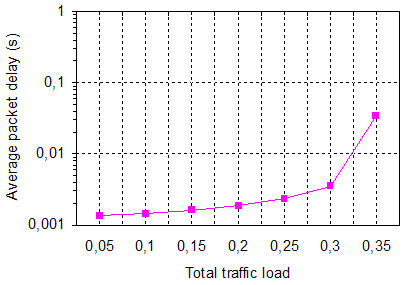}
\caption{Average packet delay of all packets}\label{results1-2}
\end{figure}

The following results include the average packet delay and the
average queue length. Each of these parameters has been evaluated
numerically from Eq. \ref{eq:packets-expected-number1}, Eq.
\ref{eq:packets-expected-number2}, and Eq.
\ref{eq:average-access-delay} by means of the formal mathematics
Maple VIII software. Fig. \ref{results1-2} shows the relationship
between average packet delay (Eq. \ref{eq:average-access-delay}) and
network traffic load (Eq. \ref{eq:throughput}). The average packet
delay is defined as the average time elapsed between the instant of
generation of a user packet and the instant of transmission of the
last bit of this packet.

In Fig. \ref{results1-2}.a (upper figure), 20 percent, 30 percent,
and 50 percent of the global offered load are EF, AF, and BE
traffic, respectively. To prevent BE queue instability, we have
observed numerically that the total traffic load must remain under
$0.4$. Hence, the offered load fluctuates from $0.05$ to $0.4$.
Under light traffic load, the results shown in Fig.
\ref{results1-2}.a can be compared to those obtained in
\cite{Luo:2005}. With an offered load of $0.4$, average packet delay
increases to almost $10$ ms. The reason is that the arriving packets
rate of BE traffic reached the service rate (i.e., $\rho_{BE} \simeq
1$).
\begin{figure}[!h]
\centering
\includegraphics[width=0.45\textwidth]{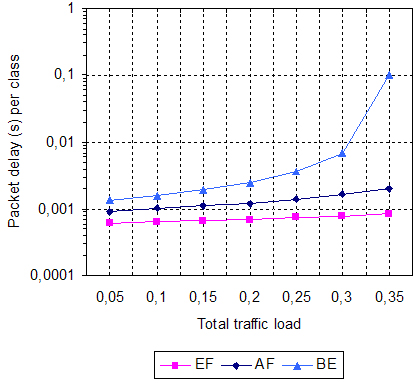}
\caption{Average packet delay per class} \label{results3}
\end{figure}
\begin{figure}
\centering
\includegraphics[width=0.45\textwidth]{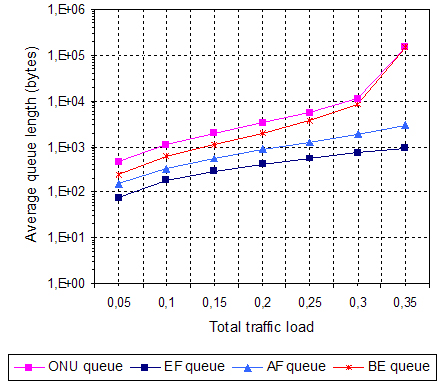}
\caption{Average queue length in bytes} \label{results4}
\end{figure}

Unlike Fig. \ref{results1-2}.a, one assumes in figure
\ref{results1-2}.b (lower figure) that the total offered load is
uniformly distributed between EF, AF, and BE traffic classes (33
percent, 33 percent, and 33 percent, respectively). The longest
average delay, observed at a load of $0.35$, is due to the BE queue
which reaches its stability limit. Figure \ref{results3} adopts the
same weight distribution between AF, EF and BE traffic classes as in
Fig. \ref{results1-2}.b. Unlike Fig. \ref{results1-2}.b, we have
plotted in Fig. \ref{results3} the average packet delay proper to
each class of traffic versus the global offered load. Finally, we
have plotted in Fig. \ref{results4} the mean number of packets in
each of the $4$ queues of our queueing model versus the offered
load, again assuming a fair distribution between the three traffic
classes. We have the confirmation of this figure that BE traffic
reaches its stability limit around a $0.35$ offered load whereas the
AF and EF traffic classes remain in their stable state. It is
possible to check the stability of the global system for an offered
load of $0.3$ by means of the Little's formula. Indeed, we see from
Fig. \ref{results4} that an average queue length of $10$e$4$ bytes
is obtained for a load equal to $0.3$ of the channel capacity. This
gives an average packet delay of $2.66$ ms. Such a delay is
confirmed by Fig. \ref{results1-2}.b.

%=================================================================
\section{Conclusion}
%=================================================================
EPONs are today considered as an economically mature alternative to
xDSL access systems. The MPCP protocol has been adopted as the IEEE
802.3ah standard MAC protocol for EPONs. Multiple dynamic bandwidth
allocation schemes such as IPACT have been recently proposed as
complement of MPCP. At the best of our knowledge, we propose in this
paper the first analytical model of the MPCP+IPACT protocol. Our
model is based on a two stages open Jackson's queueing network. Our
numerical results applied to the three Diffserv classes of traffic
are confirmed by discrete event computer simulation already
published in the recent literature.
\bibliographystyle{IEEEtran}
\bibliography{IEEEabrv,biblio}

\end{document}